\documentclass[letterpaper,twocolumn,10pt]{article}
\usepackage{usenix2019_v3}
\usepackage{graphicx}
\usepackage{placeins}
\usepackage{amsmath}
\usepackage{newtxmath}
\usepackage{xcolor}
\usepackage{booktabs}
\usepackage{multirow}
\usepackage{hyperref}
\usepackage{enumitem}
\newcommand{\name}{PrefixChain}
\title{\Large\bfseries \name: Compact Prefix Caching for Hybrid LLMs}
\author{%
\upshape Genglin Wang\textsuperscript{1}\quad Wangsong Yin\textsuperscript{2}\quad
Liekang Zeng\textsuperscript{1}\quad Yeerzhati Abudunuer\textsuperscript{1}\\
\upshape Haoxuan Xu\textsuperscript{3}\quad Zhenyu Yan\textsuperscript{1}\quad
Guoliang Xing\textsuperscript{1}\\[3pt]
\itshape\normalsize\textsuperscript{1}The Chinese University of Hong Kong\quad
\normalsize\textsuperscript{2}Peking University\\
\itshape\normalsize\textsuperscript{3}The Hong Kong University of Science and Technology
}

\begin{document}
\maketitle
\begin{abstract}
Hybrid LLMs combine linear attention with full attention to bound memory growth during long-context inference. Each linear-attention layer summarizes token history in a fixed-size recurrent state instead of a growing KV cache. However, this design creates a severe capacity bottleneck for prefix caching. Because recurrent states are overwritten in place, rolling back to prefixes requires saving bulky state checkpoints repeatedly.
We introduce \name{}, a compact prefix caching system for hybrid LLMs. Our key insight is that recurrent states evolve across tokens with strong inter-state redundancy, as each linear-attention step decays the prior state before applying an incremental update. Inspired by inter-frame video coding, \name{} organizes checkpoints into closed-loop chains of anchor I-states and low-rank residual P-states, allocating per-head ranks via offline readout-error profiling. To serve chains efficiently online, \name{} pipelines layer-grouped encoding asynchronously off the forward path, reconstructs multi-step chains in a single fused GPU kernel, and coordinates dependency-aware eviction with the KV cache. Across three hybrid LLMs and four workloads, \name{} shrinks checkpoints by $31.3\times$ to $55.7\times$ with negligible quality loss and reduces mean TTFT by up to $5.03\times$ over state-of-the-art serving systems, such as SGLang, vLLM, and Marconi, while incurring low online encoding overhead.
\end{abstract}

\section{Introduction}
Large language models (LLMs) increasingly power multi-turn dialogue, long-document analysis, and autonomous agents~\cite{openai_agents,anthropic_agents}. During inference, full-attention layers store past keys and values in a \textit{key-value (KV) cache}. When requests share a prompt prefix, serving engines widely adopt \textit{prefix caching} as a standard acceleration technique, reusing cached entries to bypass redundant prefill~\cite{zheng2024sglang,kwon2023vllm}. Because each token's KV entry depends only on preceding tokens, the cache for any shorter prefix can be retrieved directly by slicing along the sequence dimension.

However, KV caches grow linearly with sequence length and quickly dominate GPU memory during long-context serving~\cite{InfiniGen,pope2023efficiently}. To bound this memory growth, recent frontier models adopt \textit{hybrid LLM architectures}, such as Qwen3.8~\cite{qwen38_27b}, Kimi Linear~\cite{kimi_linear}, and Kimi K3~\cite{kimi_k3}. These models interleave a small number of full-attention layers with a majority of linear-attention layers. Unlike full attention, linear attention drops nonlinear softmax normalization to fold historical key-value associations into a fixed-size matrix \textit{recurrent state}~\cite{xiao2026ramnetlineartimesequencemodeling,yang2024gla,yang2024deltanet,yang2025gated}. At every token, each layer updates this state in place, by decaying the prior state and accumulating the new token's contribution. This keeps both per-token compute and active state memory $O(1)$ regardless of context length.

Unfortunately, this memory-saving design creates a severe bottleneck for prefix caching. Because recurrent states update their information in place, a sequence's final state cannot be retrieved to recover an earlier prefix (Figure~\ref{fig:attention-compare}). Therefore, enabling prefix caching requires repeatedly saving full recurrent-state \textit{checkpoints} across all linear-attention layers~\cite{sglang_unified_cache,vllm_hybrid_cache,pan2025marconi}. Meanwhile, modern workloads exhibit highly dynamic prefix hit patterns that require flexible recovery positions. For example, tree-search agents branch from arbitrary intermediate steps~\cite{yao2023tot}, and coding agents routinely edit or mask earlier tool outputs~\cite{yang2024sweagent,langchain_context_editing}. Avoiding recomputation in these workloads requires retaining many checkpoints rather than only the latest state. However, each state is huge, occupying roughly 75~MiB in Qwen3.8-27B and 222~MiB in Kimi K3, which equals thousands of full-attention KV tokens.

To cover dynamic rollback positions, a straightforward and naive approach is periodic checkpointing, which saves a full state at fixed token intervals to improve prefix hit rate. However, dense checkpoints quickly fill the memory headroom left by KV caches. To ease this pressure, production engines such as SGLang~\cite{sglang_unified_cache} save full checkpoints at request endpoints and radix-tree branch nodes under LRU eviction, while Marconi~\cite{pan2025marconi} further refines admission and eviction based on estimated compute savings. Even so, these systems share the same fundamental limitation. As contexts lengthen, growing KV caches consume most of the shared GPU memory and compete directly with recurrent states for space. Because each uncompressed checkpoint equals thousands of KV tokens, only a small number can fit under high concurrency, so many are evicted before reuse. As a result, even Marconi still recomputes nearly half of reusable tokens on SWE-agent (\S\ref{problem}). Cache policies cannot overcome this limitation alone.

Our key insight is that recurrent states along a token sequence evolve much like video frames and exhibit strong inter-state redundancy. By mathematical construction, modern linear-attention mechanisms such as GDN~\cite{yang2025gated} and KDA~\cite{kimi_linear} follow a \textit{decay-then-update} rule that scales down the prior state before adding a rank-one token update. Consequently, consecutive checkpoints retain largely the same historical context, leaving only a compact, low-rank residual between adjacent states. This residual is far more compressible than a standalone state and closely mirrors inter-frame coding in video codecs~\cite{wiegand2003h264,sullivan2012hevc}. This redundancy allows for extreme checkpoint compression, enabling limited GPU memory to accommodate far more recurrent states and fundamentally overcoming the memory capacity wall.

In this paper, we present \name{}, a compact prefix-caching system for hybrid LLMs. Borrowing I-frames and P-frames from video coding, \name{} organizes checkpoints into compressed prefix chains of self-contained \textit{I-states} and predictive \textit{P-states}. Each P-state encodes a quantized low-rank residual in a closed loop against the reconstructed predecessor, preventing errors from accumulating along the chain. Moreover, \name{} profiles head sensitivity offline to allocate heterogeneous per-head ranks under a tight memory budget. To serve these chains with low overhead, \name{} co-designs its GPU runtime across two online paths. On the \textit{checkpoint build path}, \name{} groups adjacent linear-attention layers and batches equal-rank heads across layers to avoid shape fragmentation and launch overhead. Meanwhile, an asynchronous pipeline overlaps background encoding on a separate stream with subsequent model layers. On the \textit{prefix hit path}, a single fused GPU kernel traverses the chain backward and accumulates the target state in on-chip registers, eliminating intermediate memory traffic. To manage chained records on SGLang's radix tree~\cite{zheng2024sglang}, \name{} uses dependency-aware reference counting to safely reclaim cold states without breaking active branches.

We implement \name{} atop SGLang~\cite{zheng2024sglang} and evaluate it across three hybrid LLMs on four multi-turn and agentic workloads. Experimental results show that \name{} compresses full checkpoints by $31.3\times$ to $55.7\times$ while matching BF16 accuracy up to 32K context length. Under identical GPU memory budgets, \name{} retains up to $12.1\times$ more recovery positions than Marconi and reduces mean TTFT by up to $5.03\times$ over the strongest baseline, with low encoding overhead and minor TPOT impact.
In summary, this paper makes three contributions.
\begin{itemize}[leftmargin=*, nosep, topsep=2pt, itemsep=2pt]
    \item We identify the prefix-caching capacity bottleneck in hybrid LLMs and propose closed-loop I/P-state chains with readout-guided per-head rank allocation.
    \item We build a GPU serving runtime featuring layer-grouped asynchronous encoding, single-kernel fused chain reconstruction, and dependency-aware checkpoint eviction.
    \item We evaluate \name{} across three hybrid LLMs and four workloads, achieving up to $55.7\times$ checkpoint compression and up to $5.03\times$ lower TTFT with BF16-comparable quality.
\end{itemize}

\vspace{-10pt}
\section{Background}
\label{sec:background}

\subsection{LLM Serving and Prefix Caching}\label{llm-serving-and-prefix-caching}
\noindent\textbf{LLM Serving and KV Cache.}
Decoder-only LLMs serve requests in two phases: \emph{prefill} processes the prompt and produces the first output token, while \emph{decode} generates subsequent tokens one at a time.
Full-attention layers project tokens into queries ($Q$), keys ($K$), and values ($V$) under a causal mask~\cite{vaswani2017attention}, so each token's keys and values depend only on its prefix.
The KV cache stores these representations during prefill and reuses them during decode~\cite{ye2025flashinfer}.
Serving engines use continuous batching to process concurrent requests~\cite{yu2022orca,kwon2023vllm,agrawal2024sarathi}.
Time to first token (TTFT) measures the delay from request submission to the first returned token, including queueing and prefill.
Time per output token (TPOT) measures the average interval between subsequent output tokens~\cite{zhong2024distserve}.

\begin{figure}[t]
  \centering
  \includegraphics[page=1,pagebox=cropbox,clip,width=0.9\columnwidth]{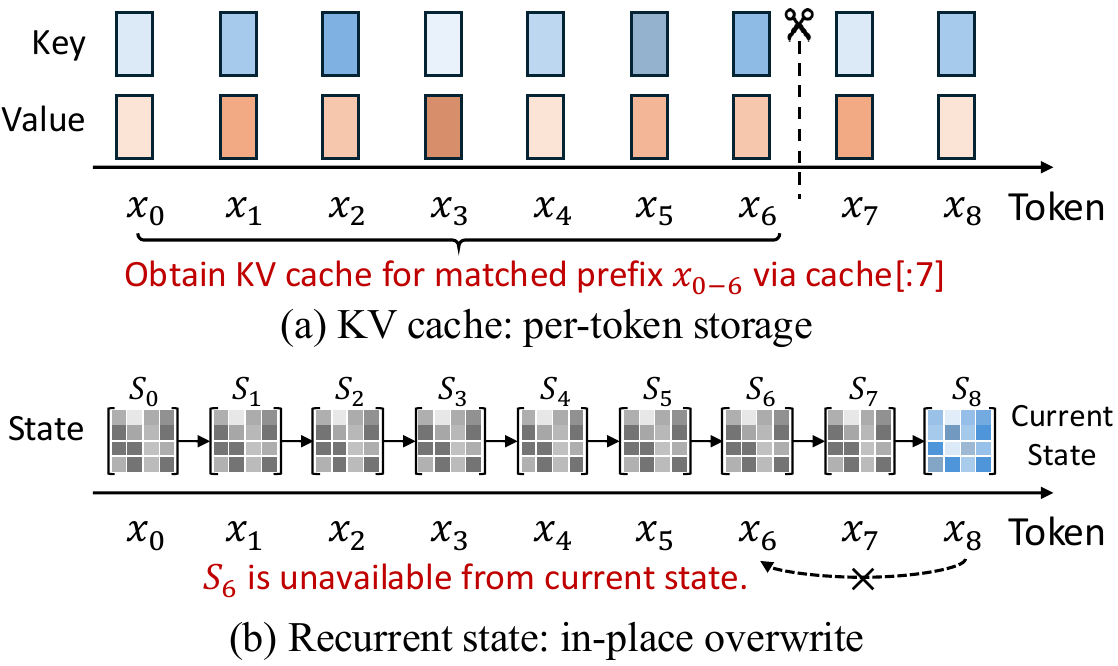}
  \vspace{-10pt}
  \caption{KV cache and recurrent state for prefix caching. (a) Token-indexed KV cache allows prefix reuse by slicing. (b) The recurrent state is overwritten in place.}
  \label{fig:attention-compare}
  \vspace{-10pt}
\end{figure}

\noindent\textbf{Prefix Caching.}
With a token-indexed KV cache, requests sharing a prompt prefix can reuse cached entries to accelerate prefill~\cite{zheng2024sglang,kwon2023vllm,gim2024promptcache}. Figure~\ref{fig:attention-compare}(a) illustrates prefix caching, where a subsequent request matching $x_{0:6}$ directly slices and reuses the corresponding KV entries. For example, SGLang's RadixAttention organizes cached prefixes in a radix tree~\cite{zheng2024sglang}, where a shared path represents common tokens and branches represent different continuations.
The engine pins entries used by active requests and evicts eligible leaf entries in least-recently-used (LRU) order.

\vspace{-10pt}
\subsection{Recurrent States in Hybrid LLMs}\label{recurrent-states-in-hybrid-llms}
\begin{figure}[t]
  \centering
  \includegraphics[page=1,pagebox=cropbox,clip,width=0.95\columnwidth]{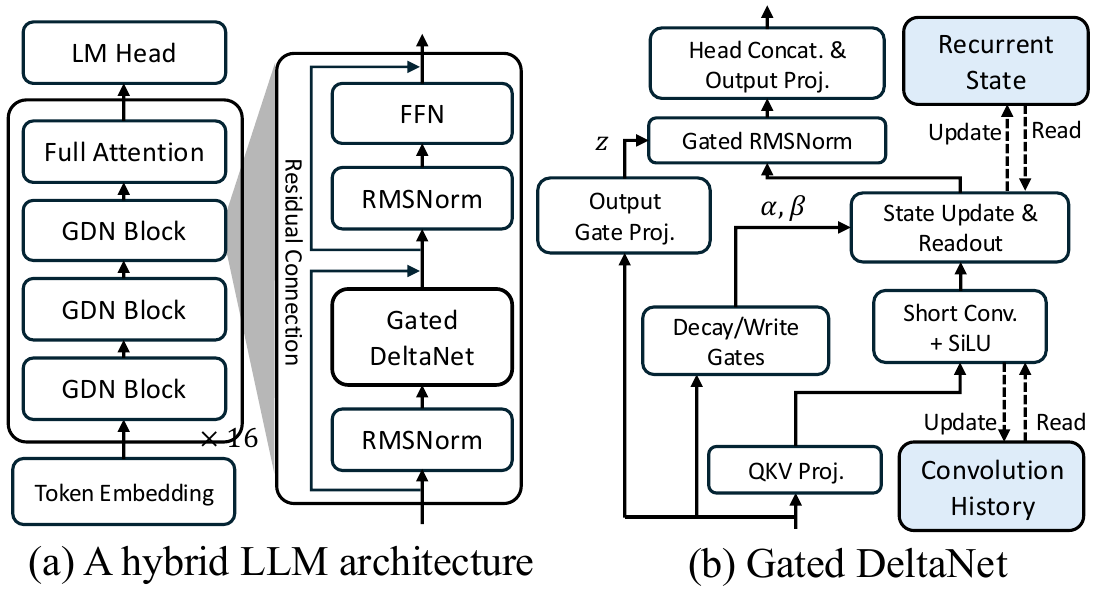}
  \vspace{-5pt}
  \caption{A hybrid LLM architecture, using Qwen3.8-27B as an example: (a) the layer stack and a GDN decoder block; (b) GDN dataflow. Q/K L2 Norm, Q/K head sharing, intermediate reshaping, and the final RMSNorm are omitted for clarity.}
  \label{fig:hybrid-architecture}
  \vspace{-15pt}
\end{figure}

\noindent\textbf{Hybrid LLMs.}
Because full-attention KV caches grow linearly with sequence length, long contexts create substantial memory pressure~\cite{InfiniGen}.
Linear attention replaces the KV cache with a fixed-size recurrent state updated at each token~\cite{katharopoulos2020transformers,schlag2021linear,yang2024gla,yang2024deltanet,yang2025gated,kimi_linear}.
However, a fixed-size state can limit accurate recall of historical details~\cite{wang2025hybrid}. To balance memory efficiency and quality, \emph{hybrid LLMs} interleave linear-attention layers with fewer full-attention layers. As shown in Figure~\ref{fig:hybrid-architecture}(a), Qwen3.8-27B~\cite{qwen38_27b} repeats three Gated DeltaNet (GDN) blocks per full-attention block.
Other frontier models, including Kimi K3~\cite{kimi_k3} and MiniMax-M1-80k~\cite{minimax_m1_config}, adopt similar hybrid designs and demonstrate strong performance on complex agentic and coding tasks.

\noindent\textbf{Recurrent States.}
Popular examples include Gated DeltaNet (GDN), used in Qwen3.8, and Kimi Delta Attention (KDA), used in Kimi Linear and Kimi K3~\cite{kimi_k3}.
Unless otherwise stated, we use GDN as the running example throughout this paper.
Figure~\ref{fig:hybrid-architecture}(b) illustrates the dataflow of a GDN module.
For one GDN head, let $S_t\in\mathbb{R}^{d_k\times d_v}$ denote the state after token $t$, where $d_k$ and $d_v$ are the key and value dimensions.
With query $q_t$, key $k_t$, value $v_t$, scalar decay gate $\alpha_t$, and update gate $\beta_t$, the state is updated as follows~\cite{yang2025gated}:
\begin{equation}\footnotesize
\begin{alignedat}{2}
\text{Decay:}\quad &\bar{S}_t &{}={}& \alpha_t S_{t-1},\\
\text{Update:}\quad &S_t &{}={}& \bar{S}_t
       +\beta_t k_t\bigl(v_t-\bar{S}_t^\top k_t\bigr)^\top,\\
\end{alignedat}
\label{eq:gdn-state}
\end{equation}
Each token step consists of \textit{Decay} and \textit{Update}.
First, the decay step scales the previous state $S_{t-1}$ by $\alpha_t$ to obtain $\bar{S}_t$. $\alpha_t$ is a scalar decay factor, controlling how much preceding information is retained.
Second, the update step adds a rank-one correction $\beta_t k_t\bigl(v_t-\bar{S}_t^\top k_t\bigr)^\top$ based on the value prediction error $v_t-\bar{S}_t^\top k_t$.
This correction adjusts the state's key-value association using the current token at $t$. Finally, query $q_t$ reads out the updated state via $y_t = S_t^\top q_t$.
These steps are applied as each token is processed during prefill and decode.

Other linear-attention variants use different decay and update rules. Gated Linear Attention (GLA) applies dimension-wise decay and adds a key-value outer product~\cite{yang2024gla}, while KDA combines dimension-wise decay with a delta-rule correction~\cite{kimi_linear}. More generally, all these variants share a \textit{decay-then-update} structure:
\begin{equation*}\footnotesize
\begin{alignedat}{2}
\text{Decay:}\quad &\bar{S}_t &{}={}& D_t \odot S_{t-1},\\
\text{Update:}\quad &S_t &{}={}& \bar{S}_t + k_t u_t^\top.
\end{alignedat}
\end{equation*}
Here $D_t$ is a scalar decay in GDN or a key-dimension-wise decay in GLA and KDA (broadcast across the value dimension). $\odot$ denotes element-wise multiplication.
The update vector is $u_t=v_t$ for GLA and $u_t=\beta_t(v_t-\bar{S}_t^\top k_t)$ for GDN and KDA. In all cases, $k_tu_t^\top$ forms a rank-one update.

Beyond the recurrent state, GDN applies width-$w$ short causal convolutions to projected queries, keys, and values for local context~\cite{yang2025gated} (Figure~\ref{fig:hybrid-architecture}(b)).
Each convolution needs the preceding $w-1$ projected input rows to process the next token.
This history is additional data needed to resume inference.
In Qwen3.8-27B ($w=4$), state recovery requires the preceding three tokens' projected QKV inputs.

\noindent\textbf{Prefix Caching for Hybrid LLMs.}
Prefix caching in hybrid LLMs must coordinate token-indexed KV caches for full-attention layers with recurrent states and convolution histories for linear-attention layers.
Because recurrent states are overwritten in place (Figure~\ref{fig:attention-compare}(b)), keeping only the final state cannot recover earlier prefixes.
Engines such as SGLang and vLLM therefore save \emph{checkpoints} snapshotting all linear-attention states and convolution histories at selected positions~\cite{sglang_unified_cache,vllm_hybrid_cache}, resuming from the nearest checkpoint on a prefix match and recomputing tokens from there onward.

\vspace{-10pt}
\section{Motivation}\label{sec:motivation}

\subsection[Problems of Prefix Caching in Hybrid LLMs]{Prefix Caching in Hybrid LLMs}\label{problem}

\begin{figure}[t]
  \centering
  \includegraphics[page=1,pagebox=cropbox,clip,width=\columnwidth]{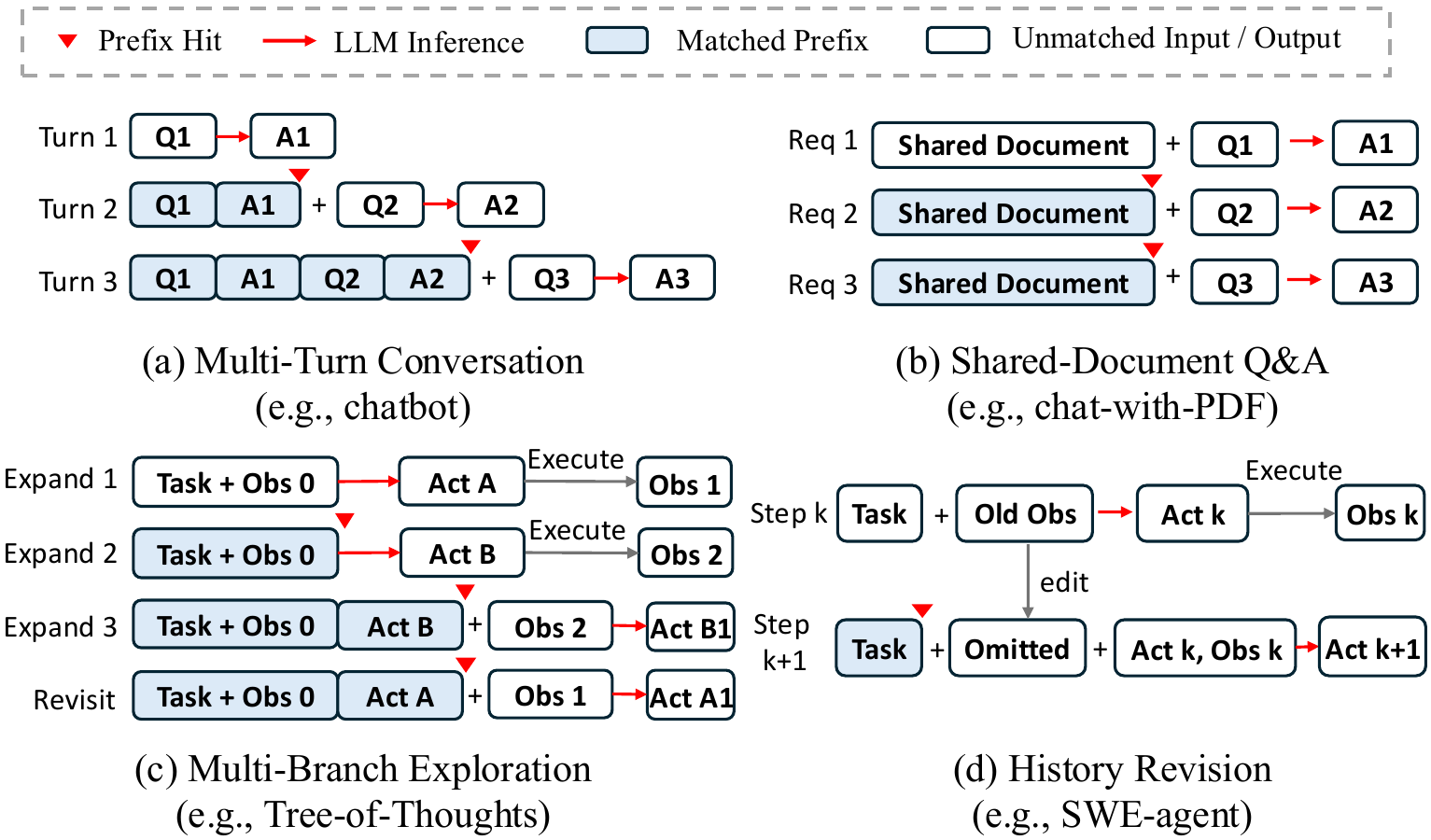}
  \vspace{-15pt}
  \caption{Dynamic prefix hit patterns in different LLM workloads, including multi-branch exploration in Tree-of-Thoughts~\cite{yao2023tot} and history revision in SWE-agent~\cite{yang2024sweagent}.}\label{fig:prefix-patterns}
  \vspace{-15pt}
\end{figure}

\noindent\textbf{Problem 1: Dynamic Prefix Hit Patterns.}
Useful prefix hit positions vary dynamically across LLM workloads, as illustrated in Figure~\ref{fig:prefix-patterns}.
Multi-turn conversations reuse the previous turn's endpoint (Figure~\ref{fig:prefix-patterns}(a)), while shared-document Q\&A repeatedly hits the document boundary across queries (Figure~\ref{fig:prefix-patterns}(b)).
Agentic workloads introduce far more complex branching and history revision.
Tree-structured search~\cite{yao2023tot,koh2025treesearch,jiang2025aide,wang2026edgecraft} explores alternative action sequences and revisits earlier nodes, creating active hit positions across multiple branches (Figure~\ref{fig:prefix-patterns}(c)).
Meanwhile, coding and tool-use agents like SWE-agent and LangChain edit or mask older tool outputs to bound context length~\cite{yang2024sweagent,langchain_context_editing}.
In Figure~\ref{fig:prefix-patterns}(d), the reusable prefix ends before the first modified token, even if later actions remain unchanged.

While KV caches support arbitrary rollback positions, keeping only the latest recurrent checkpoint leaves intermediate points uncovered.
Across 256 SWE-agent trajectories, each task accumulates 49 distinct recovery positions on average (up to 109), and concurrent serving multiplies this footprint across sessions, requiring many checkpoints to avoid recomputation.

\noindent \textbf{Problem 2: Huge Recurrent States.}
Prefix caching for hybrid LLMs places high demands on GPU memory, because each checkpoint carries a huge recurrent state. For example, one checkpoint occupies approximately 75~MiB for Qwen3.8-27B~\cite{qwen38_27b} and 222 MiB for Kimi K3~\cite{kimi_k3}, using BF16 recurrent states and convolution histories.
These sizes equal the BF16 KV cache for approximately 1.2K and 8.2K tokens, respectively. Convolution histories occupy a small fraction of each checkpoint, so recurrent states dominate checkpoint storage.

\noindent \textbf{Problem 3: High Recomputation Under Memory Pressure.}
If a matched prefix lacks a nearby checkpoint, the serving engine must recompute all uncovered prefix tokens. This increases the TTFT.
Existing systems mitigate memory pressure through cache management. SGLang evicts checkpoints and KV pages under recency~\cite{sglang_unified_cache}, while Marconi applies selective admission and FLOP-aware eviction to prioritize high-value checkpoints~\cite{pan2025marconi}. These policies improve cache utilization, but only a fraction of useful recovery positions can fit in memory because each checkpoint is still large in size.

\begin{figure}[t]
  \centering
  \includegraphics[width=\columnwidth]{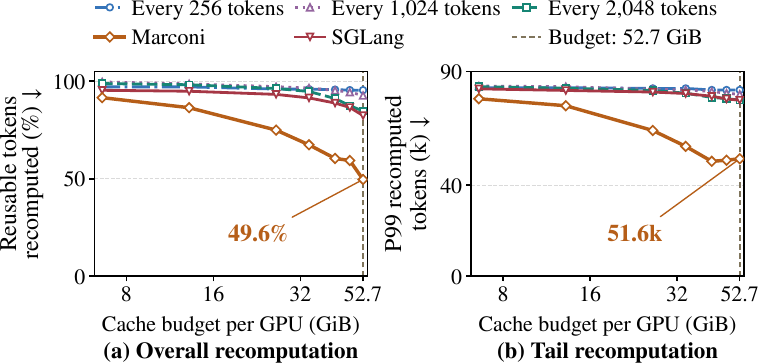}
  \caption{Prefix caching for Qwen3.8-27B on SWE-agent (up to 128 active sessions, BF16 states).
  (a) Fraction of reusable tokens recomputed.
  (b) Per-request P99 recomputed tokens.}
  \label{fig:m1-budget}
  \vspace{-10pt}
\end{figure}

Figure~\ref{fig:m1-budget} evaluates periodic BF16 checkpointing, SGLang, and Marconi on SWE-agent traces from SWE-bench Verified~\cite{chowdhury2024sweverified}.
Naive periodic BF16 checkpointing is impractical because bulky checkpoints trigger rapid eviction, forcing 95.4\% token recomputation even at the largest cache budget, so we exclude it from end-to-end serving baselines.
Selective policies still leave heavy recomputation.
At the same budget, Marconi, the strongest baseline, still recomputes 49.6\% of reusable tokens, with P99 recomputation reaching 51.6K tokens.
The root cause is capacity contention.
Because long-context KV caches consume most of the shared memory, Marconi retains an average of only 86 checkpoints across all 128 active sessions in steady state, forcing most requests to lose their latest recovery position and fall back to earlier ones.
Thus, cache policies alone cannot overcome this capacity bottleneck without compact checkpoint representations.

\vspace{-10pt}
\subsection{Compressibility of Inter-State Residual}\label{opportunity}

The problems above call for compact checkpoints. We identify three key opportunities: (1) inter-state residual compressibility, (2) heterogeneous head sensitivity, and (3) asynchronous compression off the inference critical path.

\begin{figure}[t]
  \centering
  \includegraphics[width=\columnwidth]{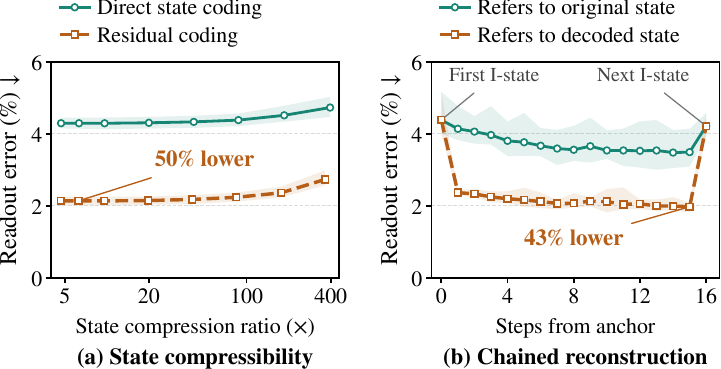}
  \vspace{-10pt}
  \caption{Recurrent-state compressibility on eight NarrativeQA documents
  with Qwen3.8-27B. (a) Residual coding lowers readout error by up to 50\%
  relative to direct coding. (b) Encoding with decoded states lowers error
  by 43\%.}
  \label{fig:m2-compressibility}
  \vspace{-10pt}
\end{figure}

\noindent\textbf{Observation 1: Inter-State Residual Compressibility.}
A naive way to shrink checkpoints is to compress each recurrent state $S_t$ independently. For example, uniform INT8 quantization reduces state size by only $2\times$, while higher compression on standalone states discards important history and degrades accuracy. However, successive recurrent states share strong temporal redundancy arising from decay and incremental updates.
Recall GDN's decay-then-update rule in Equation~\ref{eq:gdn-state}. Simplifying its rank-one update by $\Delta_t$, we have
\begin{equation*}\footnotesize
S_t = \alpha_t S_{t-1} + \Delta_t.
\end{equation*}
Here, $\Delta_t$ depends on the decayed preceding state and corrects its key-value associations.
This rule describes adjacent tokens, but two checkpoints may be separated by multiple tokens, e.g., 256 tokens at positions $s$ and $t=s+256$.
Across this interval, the state at $s$ is repeatedly decayed, and each update $\Delta_i$ is also decayed by subsequent steps.
Writing $\gamma_{a,b}=\prod_{j=a+1}^{b}\alpha_j$ for the accumulated decay, expanding these updates gives
\begin{equation}\footnotesize
S_t = \gamma_{s,t}S_s
      + \underbrace{\sum_{i=s+1}^{t}\gamma_{i,t}\Delta_i}_{R_{s,t}}.
\label{eq:gdn-checkpoint-update}
\end{equation}
Video codecs such as H.264/265 exploit inter-frame similarity by encoding residuals relative to motion-compensated reference frames~\cite{wiegand2003h264,sullivan2012hevc}.
Equation~\ref{eq:gdn-checkpoint-update} reveals a similar opportunity for residual coding in recurrent states. Given an earlier checkpoint $S_s$, instead of compressing $S_t$ from scratch, we can factor out the decayed reference $\gamma_{s,t}S_s$ and compress only the residual $R_{s,t}=S_t-\gamma_{s,t}S_s$, which isolates the incremental updates accumulated over $(s,t]$. The residual is therefore far more compressible than the full state $S_t$.

Figure~\ref{fig:m2-compressibility} validates this opportunity on NarrativeQA~\cite{kovcisky2018narrativeqa}.
As shown in Figure~\ref{fig:m2-compressibility}(a), residual coding reduces readout error (i.e., linear-attention output error on subsequent queries) by 42\% to 50\% compared to direct state coding across compression ratios.
Moreover, just as video codecs encode P-frames against \emph{reconstructed} references to prevent error drift, computing residuals relative to the decoded predecessor $\hat{S}_s$ rather than the raw state $S_s$ suppresses multi-step error accumulation by 43\% (Figure~\ref{fig:m2-compressibility}(b)).
Exploiting inter-state structure thus substantially lowers the storage cost of chained recovery positions without sacrificing accuracy.

\begin{figure}[t]
  \centering
  \includegraphics[width=0.8\columnwidth]{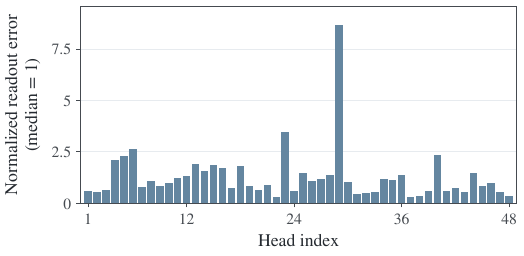}
  \vspace{-10pt}
  \caption{RMS readout error under uniform residual compression in Qwen3.8-27B (layer 32). Each bar is normalized by the median across all 48 heads.}
  \label{fig:m2-head-sensitivity}
  \vspace{-10pt}
\end{figure}

\noindent\textbf{Observation 2: Heterogeneous Head Sensitivity.}
Compressing every head's residual under the same storage budget produces highly uneven readout errors.
In linear attention, different heads track distinct context features and respond to different queries.
Because a query $q$ reads a head's recurrent state as $q^\top S$, a state error $S - \hat S$ shifts the readout by $q^\top (S - \hat S)$.
On the same eight NarrativeQA documents, we apply uniform residual compression across all heads and measure the root-mean-square readout error on subsequent queries.
Figure~\ref{fig:m2-head-sensitivity} shows all 48 heads in the 32nd linear-attention layer of Qwen3.8-27B.
The 90th-percentile error is $4.6\times$ the 10th-percentile error, and the largest error reaches $28.3\times$ the smallest.
This wide spread shows that uniform compression misallocates memory, spending equal bytes on heads with little readout impact and heads that dominate output error.
Allocating storage unevenly according to each head's readout-error reduction per byte can therefore preserve accuracy under a tighter memory budget.

\noindent\textbf{Observation 3: Asynchronous Compression Pipeline.}
Recurrent states are overwritten in place as inference advances, so preserving an intermediate recurrent state requires copying it during forward computation.
Crucially, compression itself does not lie on the current request's forward path, since the ongoing request immediately continues forward execution using the live uncompressed state.
This enables an asynchronous pipeline, in which a background encoder compresses earlier snapshots concurrently with subsequent model layers and tokens as inference advances.
Realizing this overlap without stalling inference requires carefully bounding transient snapshot buffers and arbitrating GPU compute/memory contention between model execution and encoding.

\begin{figure}[t]
  \centering
  \includegraphics[page=1,pagebox=cropbox,clip,width=0.9\columnwidth]{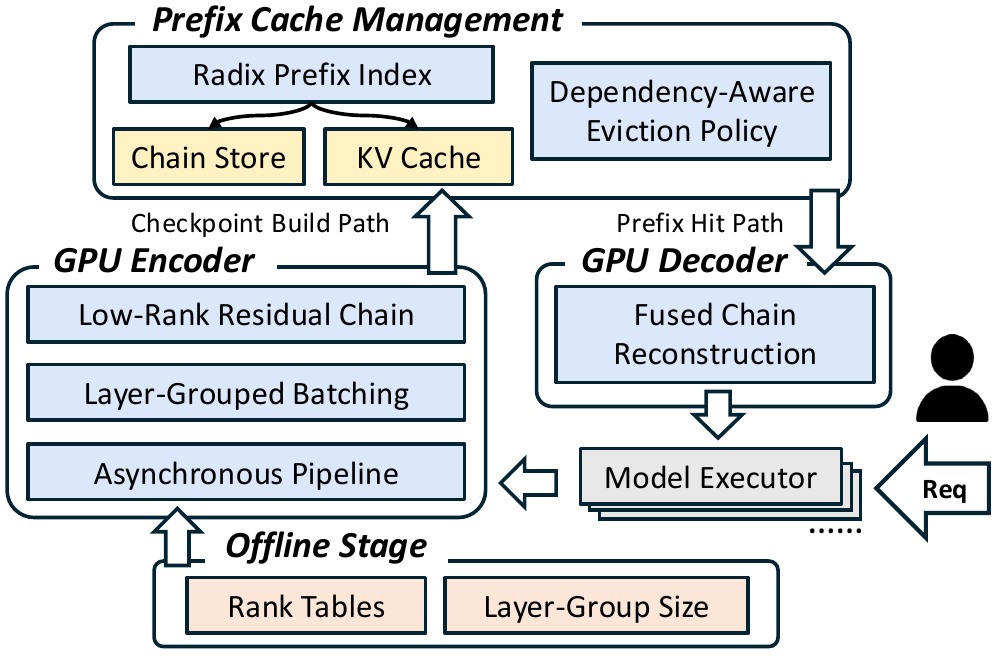}
  \vspace{-10pt}
  \caption{Overview of \name{}.}
  \label{fig:system-overview}
  \vspace{-10pt}
\end{figure}

\begin{figure*}[t]
  \centering
  \includegraphics[page=1,pagebox=cropbox,clip,width=\textwidth]{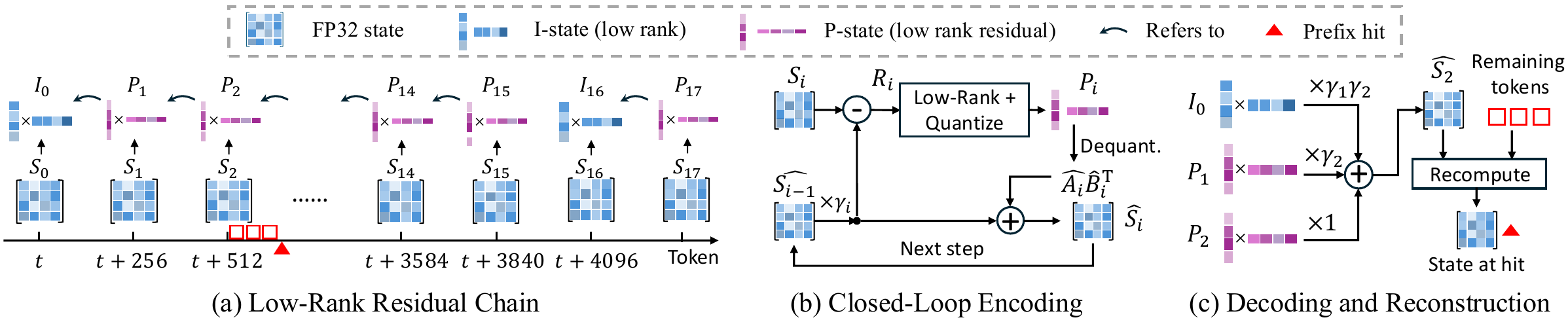}
  \vspace{-13pt}
  \caption{Overview of Low-Rank Residual Chain. (a) This example stores one I-state followed by 15 residual P-states. (b) Each P-state encodes a residual using the previous reconstructed state as a reference. (c) On a prefix hit position, \name{} reconstructs the state at the nearest preceding recovery point and recomputes the remaining tokens.}
  \label{fig:chained}
  \vspace{-10pt}

\end{figure*}

\vspace{-10pt}
\section{Overview of \name{}}
\label{sec:overview}

Figure~\ref{fig:system-overview} shows the architecture of \name{}, a prefix-caching system that compresses recurrent-state checkpoints. Inspired by inter-frame video coding in H.264/H.265~\cite{wiegand2003h264,sullivan2012hevc}, \name{} views recurrent states along a sequence as video frames and organizes checkpoints into chains of \textit{I-states} and \textit{P-states}. Like a video keyframe (I-frame), each chain begins with a self-contained I-state anchor to prevent error accumulation, while subsequent P-states act as predictive P-frames that record only compact incremental updates from the preceding reconstructed state, amortizing storage across successive prefix positions. Around the \textit{Model Executor}, \name{} operates through an offline stage and two online serving paths:
\begin{itemize}[leftmargin=*, nosep, topsep=3pt, itemsep=2pt]
    \item \textbf{Offline Stage.} Before deployment, \name{} calibrates fixed per-head rank tables under a storage budget (\S\ref{sec:per-head-rank-allocation}) and selects the layer-group size for the target model and GPU platform (\S\ref{sec:gpu-encoding-decoding}).
    \item \textbf{Checkpoint Build Path.} During inference, the model executor captures states at selected boundaries and passes them to the \textit{GPU Encoder}. Using layer-grouped batching and an asynchronous pipeline, the encoder packs states into low-rank residual chains, stores completed records in the \textit{Chain Store}, and links them with \textit{KV Cache} blocks in the \textit{Radix Prefix Index}.
    \item \textbf{Prefix Hit Path and Cache Management.} For each incoming request, \textit{Prefix Cache Management} finds the deepest valid match in the radix index. On a hit, the \textit{GPU Decoder} reconstructs the target state from the chain store so the model executor processes only the remaining suffix. A dependency-aware eviction policy coordinates reclamation across the chain store and KV cache (\S\ref{sec:cache-mgmt}).
\end{itemize}

\vspace{-10pt}
\section{Compressed Prefix Chain}\label{sec:compressed-prefix-chain}

\subsection{Low-Rank Residual Chain}
\label{sec:low-rank-residual-coding}

\noindent\textbf{Chained Checkpoints.}
\name{} exploits the inter-state compressibility from \S\ref{opportunity} by organizing periodic FP32 checkpoints into a compressed chain (Figure~\ref{fig:chained}(a)).
For a given layer and head, let $S_i \in \mathbb{R}^{d_k \times d_v}$ denote the exact state at checkpoint $i$ and $\hat S_i$ denote its reconstruction.
As illustrated in Figure~\ref{fig:chained}(a) with 256-token intervals, an anchor I-state $I_0$ independently compresses $S_0$, and each subsequent P-state $P_1, \ldots, P_{15}$ encodes only the state residual relative to its preceding reconstruction $\hat S_{i-1}$.
Multiple recovery positions thus share the anchor and earlier P-states.
At checkpoint $16$ ($t+4096$), $I_{16}$ starts a new chain to bound the maximum reconstruction depth.

\noindent\textbf{Closed-Loop Low-Rank Decomposition.}
Figure~\ref{fig:chained}(b) shows how \name{} encodes step $i$.
In GDN, the accumulated scalar decay $\gamma_i$ over the checkpoint interval scales $\hat S_{i-1}$, and the encoder subtracts this decayed reference from $S_i$ to approximate the residual $R_i$ with a rank-$r$ decomposition
\begin{equation*}\footnotesize
R_i = S_i - \gamma_i \hat S_{i-1}
    \approx A_i B_i^\top,
\end{equation*}
where $A_i \in \mathbb{R}^{d_k \times r}$, $B_i \in \mathbb{R}^{d_v \times r}$, and $r$ is the head-specific rank (\S\ref{sec:per-head-rank-allocation}).
Quantizing $A_i$ and $B_i$ to INT4 yields the compact record $P_i$, which stores $r(d_k + d_v)$ quantized elements alongside quantization scales and $\gamma_i$ instead of $d_k d_v$ entries.
For KDA, we store one accumulated decay factor per key dimension and apply the same procedure. I-states apply the same low-rank quantization directly to $S_0$.

Inspired by closed-loop video coding~\cite{wiegand2011source}, \name{} locally dequantizes $P_i$ into $\hat R_i = \hat A_i \hat B_i^\top$ to keep encoder and decoder references identical.
It then reconstructs the reference for the next step via
\begin{equation}\footnotesize
\hat S_i = \gamma_i \hat S_{i-1} + \hat R_i.
\label{eq:chain-reconstruction}
\end{equation}
An open-loop encoder would instead reference the exact $S_{i-1}$ while the decoder only has $\hat S_{i-1}$, which causes prediction drift across steps.
Referencing $\hat S_{i-1}$ absorbs prior approximation errors into the current residual and ensures $S_i - \hat S_i = R_i - \hat R_i$.
In Figure~\ref{fig:m2-compressibility}(b), this closed-loop design cuts step-15 readout error by 43\% over open-loop coding.

\noindent\textbf{Decoding and Reconstruction.}
Figure~\ref{fig:chained}(c) shows how \name{} serves the prefix hit marked by the red triangle after $t+512$ in Figure~\ref{fig:chained}(a).
To recover the nearest preceding checkpoint $i$, \name{} dequantizes and multiplies the stored factors of $I_0$ and $P_1, \ldots, P_i$.
This step yields $\hat S_0$ and $\hat R_1, \ldots, \hat R_i$.
Unrolling Equation~\ref{eq:chain-reconstruction} gives the target state
\begin{equation}\footnotesize
\hat S_i =
\left(\prod_{j=1}^{i}\gamma_j\right)\hat S_0
+ \sum_{k=1}^{i}
\left(\prod_{j=k+1}^{i}\gamma_j\right)\hat R_k,
\label{eq:chain-decoding}
\end{equation}
where an empty product equals $1$.
In Figure~\ref{fig:chained}(c), checkpoint $2$ weights $I_0$, $P_1$, and $P_2$ by $\gamma_1\gamma_2$, $\gamma_2$, and $1$ to form $\hat S_2 = \gamma_1\gamma_2 \hat S_0 + \gamma_2 \hat R_1 + \hat R_2$.
\name{} then resumes inference from $\hat S_2$ and recomputes the remaining shared tokens to obtain the state at the hit position.

\vspace{-10pt}
\subsection{Offline Per-Head Rank Allocation}
\label{sec:per-head-rank-allocation}

Under uniform compression, readout error varies widely across heads because different heads track distinct context features and respond differently to queries (\S\ref{opportunity}, Figure~\ref{fig:m2-head-sensitivity}).
Moreover, since a state error $S - \hat S$ perturbs linear-attention outputs by $q^\top(S - \hat S)$, equal-norm state errors can affect readouts very differently across query directions.
To preserve accuracy under a fixed byte budget, \name{} allocates heterogeneous per-head ranks offline instead of compressing every head uniformly, producing two fixed rank tables across layers and heads for I-states and P-states.

Specifically, \name{} profiles I-states on exact states $S_0$ and P-states on closed-loop residuals $R_i$ generated under fixed reference ranks (\S\ref{sec:low-rank-residual-coding}) over a calibration corpus.
For each head and candidate rank, it measures the mean-squared readout error on calibration queries and averages across samples to build per-head rank-error curves.
Building on sensitivity-aware rank allocation~\cite{chang2025palu}, a greedy solver then assigns ranks by readout-error reduction per additional byte.
For a chain of one I-state and 15 P-states, it weights I-state and P-state error and storage bytes (including quantization scales) by $1/16$ and $15/16$.
Starting from the minimum permitted ranks, the solver iteratively picks the rank increment with the largest error reduction per byte until no feasible increment fits within the budget.
The resulting rank tables remain fixed across requests, keeping rank search off the online serving path.

\vspace{-10pt}
\section{GPU Encoding and Decoding}
\label{sec:gpu-encoding-decoding}

\begin{figure}[t]
  \centering
  \includegraphics[page=1,pagebox=cropbox,clip,width=\columnwidth]{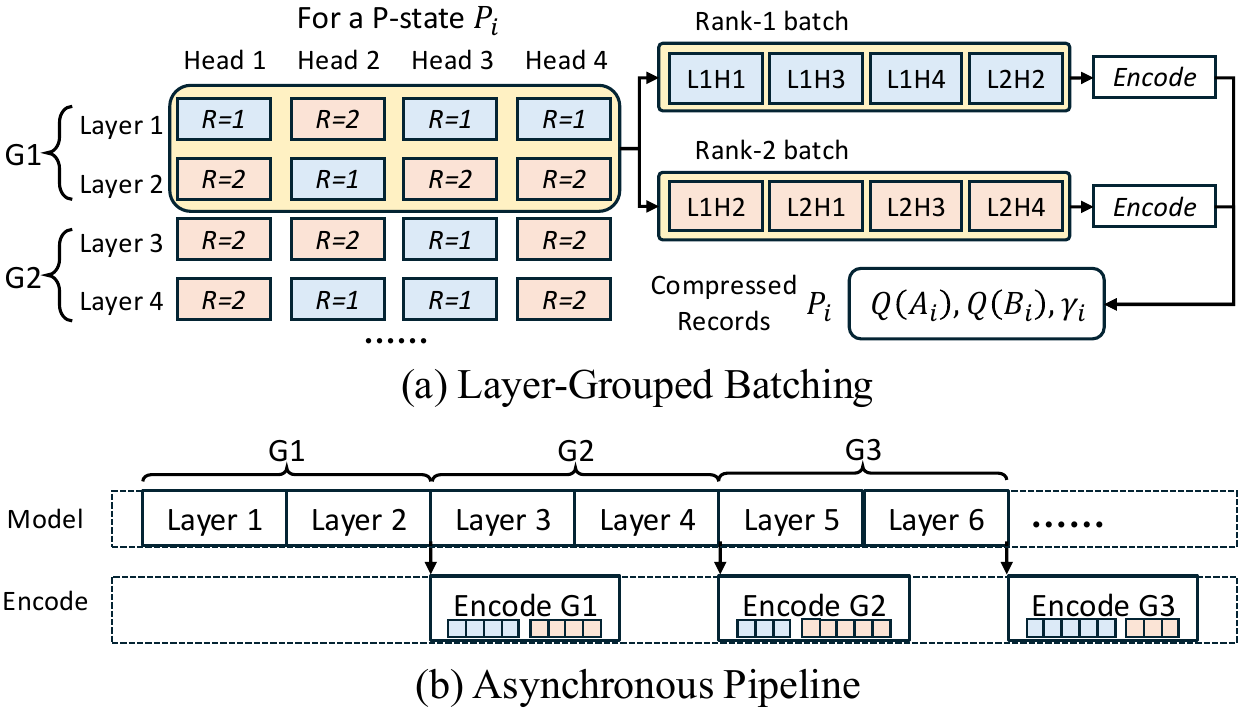}
  \vspace{-10pt}
  \caption{GPU encoding. This example groups two linear-attention layers. Full-attention layers are omitted.}
  \label{fig:gpu-encoding}
  \vspace{-10pt}
\end{figure}

\vspace{-10pt}
\subsection{Checkpoint Build Path}

\noindent\textbf{Layer-Grouped Batching.}
Encoding checkpoints naively suffers from heavy kernel launch overhead and low GPU utilization due to two sources of fragmentation.
First, per-head rank allocation splits each layer into small subsets of different rank shapes, preventing heads with different ranks from sharing a single batched decomposition kernel without wasteful padding.
Second, encoding one layer at a time launches small workloads across dozens of layers and underutilizes the GPU.

\begin{figure}[t]
  \centering
  \includegraphics[width=0.95\columnwidth]{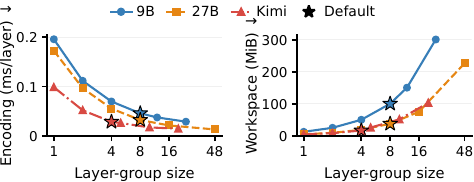}
  \vspace{-5pt}
  \caption{Per-layer encoding time and workspace memory across layer-group sizes on an 80\,GiB A800 GPU (TP1 for Qwen3.5-9B and TP4 for Qwen3.8-27B and Kimi-Linear). Stars mark our default group sizes of 8, 8, and 4.}
  \label{fig:layer-group-size}
  \vspace{-15pt}
\end{figure}

To eliminate both forms of fragmentation, \name{} combines layer grouping with rank-based batching. In Figure~\ref{fig:gpu-encoding}(a), it first groups adjacent linear-attention layers into a layer group, such as G1 spanning Layers 1 and 2, pooling more heads into one encoding unit to reduce per-layer launch overhead.
After grouping layers, \name{} merges heads with the same assigned rank across those layers into uniform-shape batches.
In Figure~\ref{fig:gpu-encoding}(a), the four rank-1 heads across both layers form one batch, and the four rank-2 heads form another.
Rank-specialized kernels then encode each batch in parallel via subspace iteration to produce $P_i$, while preserving sequential closed-loop order against $\hat S_{i-1}$ (\S\ref{sec:low-rank-residual-coding}).
Before deployment, we profile encoding latency and workspace memory across layer-group sizes (Figure~\ref{fig:layer-group-size}).
Per-layer latency drops sharply before flattening, while workspace memory grows linearly.
Picking defaults at this knee (8 layers for both Qwen models and 4 for Kimi-Linear) captures most of the speedup without inflating memory.

\noindent\textbf{Asynchronous Pipeline.}
Ongoing inference of each layer group advances from its live uncompressed state (\S\ref{opportunity}).
Capturing a state at fixed token intervals is lightweight in both serving phases.
During prefill, serving engines run chunkwise linear-attention kernels derived from Flash Linear Attention (FLA)~\cite{flash_linear_attention}, which process prompts chunk by chunk and already materialize the recurrent state at every chunk boundary.
Aligning the checkpoint interval with a multiple of the FLA chunk size lets \name{} copy boundary states directly without extra attention passes.
During decode, the recurrent state updates in place at each token, so \name{} copies the live state immediately upon reaching a checkpoint position.

\name{} places these copied states into a preallocated, bounded GPU staging pool and runs the encoder asynchronously on a separate CUDA stream.
As illustrated in Figure~\ref{fig:gpu-encoding}(b), once the forward pass finishes layer group G1, the encoder waits on a capture event and encodes G1 while the model executes the next group G2.
Co-scheduling works on the same GPU because forward operators do not uniformly saturate all compute and memory resources, especially during memory-bound decode steps, linear-attention kernels, and inter-kernel gaps.
Moreover, grouped encoding takes well under a millisecond per group (Figure~\ref{fig:layer-group-size}), which is shorter than the forward pass of the next layer group and its interleaved full-attention layers.
The background encoder can therefore absorb idle GPU capacity and keep pace with inference without stalling the forward stream.

\noindent\textbf{Entropy Coding.}
Even after INT4 quantization, the packed factors follow skewed distributions with substantial redundancy.
\name{} applies lossless GPU asymmetric numeral systems (ANS) coding to each packed record ($Q(A_i)$, $Q(B_i)$, FP32 scales, and $\gamma_i$) before storing it, further shrinking cache footprint without adding error.
On a prefix hit, parallel GPU ANS decoding unpacks the needed records directly.

\vspace{-10pt}
\subsection{Prefix Hit Path}
\label{sec:prefix-hit-path}

\begin{figure}[t]
  \centering
  \includegraphics[page=1,pagebox=cropbox,clip,width=\columnwidth]{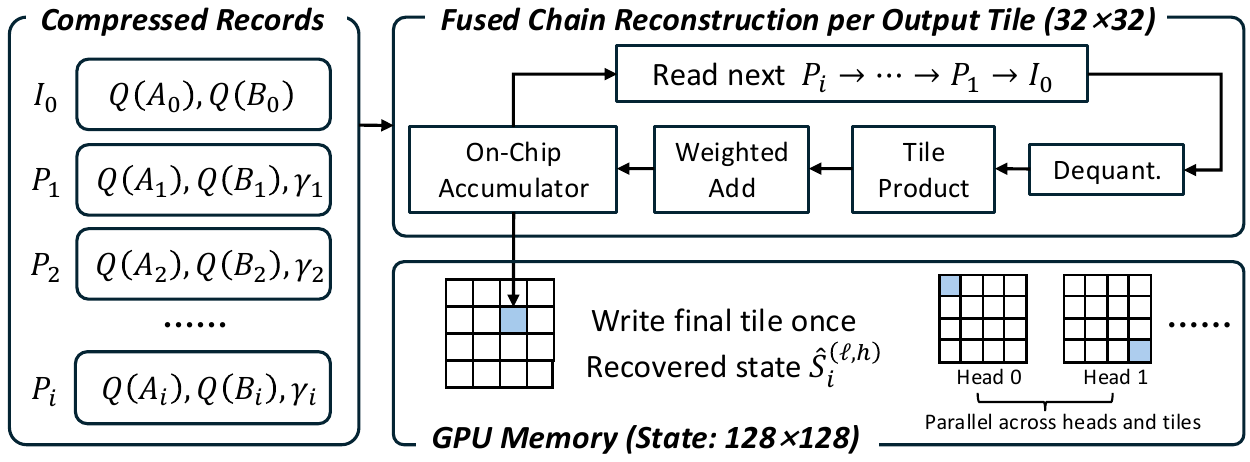}
  \vspace{-15pt}
  \caption{Fused chain reconstruction.}
  \label{fig:fused-chain-reconstruction}
  \vspace{-15pt}
\end{figure}

\noindent\textbf{Fused Chain Reconstruction.}
After GPU ANS decoding, reconstructing a target checkpoint may depend on an anchor I-state and up to 15 P-states.
Materializing a full state at each chain step would repeatedly write and reread complete matrices in GPU memory.
Because Equation~\ref{eq:chain-decoding} expresses $\hat S_i$ as a decay-weighted sum of low-rank terms, \name{} traverses the chain backward from $P_i$ to $I_0$ in a single fused kernel (Figure~\ref{fig:fused-chain-reconstruction}).
It partitions each $128 \times 128$ head state into $32 \times 32$ output tiles and launches a 3D GPU grid across all layers, heads, and tiles in parallel.
Because every tile program owns a disjoint output block, all 16 tiles per head execute concurrently without cross-block synchronization or atomic updates.
Each program for tile $(m, n)$ keeps an on-chip FP32 accumulator $\mathrm{acc} \in \mathbb{R}^{32 \times 32}$ initialized to zero and a running decay weight $\alpha$ initialized to $1$.
For $k = i, i-1, \ldots, 0$, it updates both states as
\begin{equation*}\footnotesize
\mathrm{acc} \leftarrow \mathrm{acc} + (\alpha \hat A_k^{(m)})(\hat B_k^{(n)})^\top,
\qquad
\alpha \leftarrow \alpha \gamma_k,
\end{equation*}
where $\hat A_k^{(m)}$ and $\hat B_k^{(n)}$ are the 32-row slices of $\hat A_k$ and $\hat B_k$ for tile $(m, n)$.
Accumulating record $k$ before multiplying $\alpha$ by $\gamma_k$ ensures $\gamma_k$ decays only earlier states and avoids rescaling $\mathrm{acc}$ across steps.
In KDA layers, $\alpha$ and $\gamma_k$ extend to 32-element row vectors without changing the grid or loop.
On chip, the kernel unpacks only the needed INT4 slices and folds $\alpha$ into the left-factor scale during multiply-accumulate, avoiding intermediate tile buffers.
After accumulating $I_0$, it writes the finished tile of $\hat S_i^{(\ell,h)}$ once directly into the engine's active state slot, eliminating intermediate state traffic and per-step kernel launches.

\noindent\textbf{Short-Convolution History Reconstruction.}
Resuming inference also requires restoring each linear layer's short causal convolution history.
For convolution width $w$, \name{} saves only the $w-1$ shared BF16 hidden vectors entering the Q, K, and V projections at the checkpoint boundary (three vectors per layer in Qwen3.8-27B with $w=4$).
On recovery, it replays those projections with the model's existing weights to repopulate the larger Q, K, and V convolution buffers, trading minimal replay compute for smaller persistent storage.

\vspace{-10pt}
\subsection{Prefix Cache Management}
\label{sec:cache-mgmt}

Serving compressed chains efficiently requires managing state records and short-convolution histories alongside the KV cache.
\name{} does not reinvent prefix cache management.
Instead, it attaches compressed checkpoints directly to SGLang's radix tree~\cite{zheng2024sglang} while storing their bytes in a separate GPU chain store (Figure~\ref{fig:system-overview}).
A prefix hit requires three matching pieces at the recovery point, namely the KV cache, the recurrent state, and its short-convolution history.
On arrival, a request resumes from the deepest checkpoint where all three are valid and recomputes any remaining suffix.

\noindent\textbf{Dependency-Aware Eviction.}
Uncompressed engines can evict any earlier state while keeping later ones valid, because each state summarizes all prior tokens~\cite{sglang_unified_cache}.
In \name{}, eviction depends on chain structure across three cases.
(1)~\textit{Across I/P chains.} Periodic I-states along the same token sequence start independent chains with no cross-chain dependency, so an earlier chain can be evicted freely without affecting later ones.
(2)~\textit{Within an I/P chain.} Inside a chain ($I_0 \to P_1 \to P_2 \to \cdots$), \name{} adds two-stage reference-counted eviction atop SGLang's LRU.
If $P_1$ turns cold while $P_2$ remains active, \name{} removes $P_1$ from matching and immediately frees its short-convolution history, which dominates compressed checkpoint bytes on Qwen (\S\ref{sec:eval-compression}), because $P_2$ only needs its own history.
Meanwhile, $P_1$'s compressed state record stays resident until downstream checkpoints and active reconstructions release their references, after which reclamation cascades to unreferenced parents.
(3)~\textit{Branching inside an I/P chain.} When requests diverge from inside a chain on the radix tree, the chain naturally forms a tree where child branches point to shared parent records without copying.
Reference counting keeps shared ancestors resident until every child branch releases them, while letting each branch's own P-states be evicted independently.
Finally, eviction between KV blocks and checkpoints on the same tree is asymmetric.
Dropping KV blocks invalidates every checkpoint relying on those tokens, whereas evicting a checkpoint leaves KV blocks intact for deeper checkpoints or sibling branches.
We use LRU to keep the focus on compression, while value-based policies such as Marconi~\cite{pan2025marconi} are orthogonal and can sit atop the same mechanism.

\vspace{-10pt}
\section{Implementation}
\label{sec:implementation}

\noindent\textbf{SGLang Integration and GPU Runtime.}
We build \name{} on SGLang in roughly 6.4K lines of Python for cache management and 0.5K lines of Triton for GPU codec kernels.\footnote{AI use disclosure: AI coding agents assisted with parts of the implementation, evaluation scripts, and writing refinement under author guidance. All designs, insights, and conclusions are by the authors.}
The cache manager hooks into SGLang's radix tree and KV eviction path to enforce asymmetric invalidation without altering the attention scheduler, while storing compressed I/P records and convolution inputs in a separate bounded GPU byte arena.
Because offline rank tables fix batch shapes, \name{} captures each layer group's encoding pipeline into reusable buffers and CUDA Graphs, and runs GPU ANS coding on device buffers via NVIDIA nvCOMP.
Each active chain-building request holds one temporary FP32 reference state (80 to 192\,MiB peak per GPU, \S\ref{sec:eval-overhead}), independent of history length, and releases it once encoding finishes so later extensions rebuild from cache.

\noindent\textbf{Distributed Inference.}
\name{} supports tensor-parallel (TP) inference along the model's head-sharded layout.
Each GPU stores rank tables, states, and compressed records for its local heads, running encoding, ANS coding, and fused reconstruction locally without transferring state matrices.
To avoid duplicating convolution inputs, each rank stores a feature shard of the saved hidden vectors, all-gathers them on recovery, and replays local QKV projections.
At scheduler barriers, ranks exchange small checkpoint IDs and readiness flags to expose checkpoints only after all ranks finish encoding and keep eviction and recovery synchronized with negligible overhead.
Because coding and convolution recovery are layer-local, this design extends directly to pipeline parallelism without cross-stage state transfer.

\vspace{-10pt}
\section{Evaluation}
\label{sec:evaluation}

\vspace{-10pt}
\subsection{Experimental Setup}
\label{sec:eval-setup}

\noindent\textbf{Models and Testbed.}
We evaluate three hybrid LLMs covering both linear-attention families, namely Qwen3.5-9B~\cite{qwen35_9b} and Qwen3.8-27B~\cite{qwen38_27b} (GDN) and Kimi-Linear-48B-A3B~\cite{kimi_linear} (KDA).
All experiments run on a server with four NVLink-connected NVIDIA A800-SXM4 GPUs (80\,GiB each) and two 32-core Intel Xeon Platinum 8378A CPUs, using CUDA 12.6, PyTorch 2.13.0, Triton 3.7.1, and NVIDIA nvCOMP 5.3.0.
Qwen3.5-9B runs on one GPU, while Qwen3.8-27B and Kimi-Linear run with four-way tensor parallelism.

\begin{table}[t]
  \centering
  \fontsize{9}{10.5}\selectfont
  \setlength{\tabcolsep}{3.5pt}
  \renewcommand{\arraystretch}{1.08}
  \caption{Serving workloads.}
  \begin{tabular}{@{}llrr@{}}
    \toprule[0.6pt]
    & & \multicolumn{2}{c}{\textbf{Requests}} \\
    \cmidrule(lr){3-4}
    \textbf{Workload} & \textbf{Scale} & \textbf{Cache build} & \textbf{Timed visits} \\
    \midrule[0.4pt]
    ShareGPT~\cite{sharegpt_dataset} & 2,048 sessions & 5,976  & 512 \\
    ToolMind~\cite{yang2025toolmind} & 1,024 sessions & 3,933  & 512 \\
    LooGLE~\cite{li2024loogle} & 140 docs       & 140    & 140 \\
    SWE-agent~\cite{yang2024sweagent} & 128 tasks      & 5,915  & 128 \\
    \bottomrule[0.6pt]
  \end{tabular}
  \vspace{-5pt}
  \label{tab:serving-workloads}
  \vspace{-5pt}
\end{table}

\noindent\textbf{Configurations.}
Model weights, KV caches, and active recurrent states stay in BF16.
\name{} places checkpoints every 256 tokens and anchor I-states every 4,096 tokens, using one fixed offline rank table per model across all benchmarks without per-task tuning. All compared systems share the same GPU allocation and per-GPU memory limit.
Offline per-head rank calibration uses eight 8,192-token segments from the WikiText-2 training split, disjoint from the WikiText-2 test split used for PPL evaluation.

\noindent\textbf{Baselines.}
For compression quality, BF16 states serve as the lossless reference alongside uniform INT8 quantization.
For serving, we compare \name{} against SGLang~\cite{zheng2024sglang} (v0.5.21), vLLM~\cite{kwon2023vllm} (v0.31.0), and Marconi~\cite{pan2025marconi}, all with hybrid prefix caching enabled.
We integrate Marconi directly into SGLang to remove cross-engine differences.

\begin{table*}[t]
  \centering
  \fontsize{8.8}{10}\selectfont
  \setlength{\tabcolsep}{2.2pt}
  \renewcommand{\arraystretch}{0.5}
  \begin{tabular}{@{}lllc@{\hspace{5pt}}c@{\hspace{4pt}}c@{\hspace{4pt}}c@{\hspace{5pt}{\color{black!25}\vrule width 0.3pt}\hspace{5pt}}c@{\hspace{4pt}}c@{\hspace{4pt}}c@{\hspace{5pt}{\color{black!25}\vrule width 0.3pt}\hspace{5pt}}c@{\hspace{4pt}}c@{\hspace{4pt}}c@{}}
    \toprule[0.6pt]
    & & & & \multicolumn{3}{c}{\textbf{Qwen3.5-9B}} & \multicolumn{3}{c}{\textbf{Qwen3.8-27B}} & \multicolumn{3}{c}{\textbf{Kimi-Linear}} \\
    \cmidrule(l{2pt}r{6pt}){5-7}\cmidrule(l{6pt}r{6pt}){8-10}\cmidrule(l{6pt}r{2pt}){11-13}
    \textbf{Benchmark} & \textbf{Task} & \textbf{Metric} & \textbf{Length}
    & BF16 & INT8 & \name{} & BF16 & INT8 & \name{} & BF16 & INT8 & \name{} \\
    \midrule[0.4pt]
\multirow{5}{*}{\textit{LongBench}}   &   NarrativeQA   &   F1 $\uparrow$   &   $\leq$32K   & 30.8 & 30.5 & 30.0 & 34.5 & 34.6 & 34.1 & 33.7 & 32.0 & 32.1 \\
    &   Qasper   &   F1 $\uparrow$   &   $\leq$32K   & 51.4 & 51.4 & 50.4 & 48.2 & 48.5 & 49.0 & 47.2 & 47.3 & 46.9 \\
    &   HotpotQA   &   F1 $\uparrow$   &   $\leq$32K   & 61.3 & 62.0 & 60.7 & 70.2 & 70.2 & 70.6 & 62.3 & 62.9 & 61.9 \\
    &   GovReport   &   ROUGE-L $\uparrow$   &   $\leq$32K   & 31.2 & 31.1 & 30.8 & 33.2 & 33.0 & 33.1 & 30.9 & 30.9 & 31.1 \\
    &   RepoBench-P   &   Edit similarity $\uparrow$   &   $\leq$32K   & 61.1 & 60.3 & 62.1 & 39.7 & 39.2 & 38.5 & 64.6 & 65.2 & 65.2 \\
    \midrule[0.4pt]
\multirow{12}{*}[-8pt]{\textit{RULER}}   &   \multirow{3}{*}{\shortstack[l]{Multi-key\\retrieval}}   &   \multirow{3}{*}{Match score $\uparrow$}   &   8K   & 100.0 & 100.0 & 100.0 & 100.0 & 100.0 & 100.0 & 100.0 & 100.0 & 100.0 \\
    &      &      &   16K   & 100.0 & 100.0 & 100.0 & 100.0 & 100.0 & 100.0 & 100.0 & 100.0 & 100.0 \\
    &      &      &   32K   & 100.0 & 100.0 & 100.0 & 100.0 & 100.0 & 100.0 & 100.0 & 100.0 & 100.0 \\
    \addlinespace[1pt]
    \cmidrule[0.25pt]{2-13}
    \addlinespace[1pt]
    &   \multirow{3}{*}{\shortstack[l]{Multi-value\\retrieval}}   &   \multirow{3}{*}{Match score $\uparrow$}   &   8K   & 100.0 & 100.0 & 98.2 & 100.0 & 100.0 & 100.0 & 99.8 & 99.8 & 99.5 \\
    &      &      &   16K   & 100.0 & 99.5 & 99.5 & 100.0 & 100.0 & 100.0 & 99.5 & 99.8 & 98.5 \\
    &      &      &   32K   & 99.8 & 99.0 & 99.8 & 100.0 & 100.0 & 100.0 & 98.2 & 98.2 & 96.8 \\
    \addlinespace[1pt]
    \cmidrule[0.25pt]{2-13}
    \addlinespace[1pt]
    &   \multirow{3}{*}{Variable tracking}   &   \multirow{3}{*}{Match score $\uparrow$}   &   8K   & 100.0 & 100.0 & 100.0 & 100.0 & 100.0 & 100.0 & 99.8 & 100.0 & 99.8 \\
    &      &      &   16K   & 100.0 & 100.0 & 100.0 & 100.0 & 100.0 & 100.0 & 99.8 & 99.8 & 99.8 \\
    &      &      &   32K   & 100.0 & 100.0 & 100.0 & 100.0 & 100.0 & 100.0 & 99.6 & 99.6 & 99.6 \\
    \addlinespace[1pt]
    \cmidrule[0.25pt]{2-13}
    \addlinespace[1pt]
   &   \multirow{3}{*}{Common words}   &   \multirow{3}{*}{Match score $\uparrow$}   &   8K   & 100.0 & 100.0 & 100.0 & 100.0 & 100.0 & 100.0 & 97.8 & 96.9 & 96.9 \\
   &      &      &   16K   & 99.2 & 99.0 & 100.0 & 100.0 & 100.0 & 100.0 & 98.1 & 97.7 & 97.8 \\
   &      &      &   32K   & 100.0 & 100.0 & 100.0 & 100.0 & 100.0 & 100.0 & 97.3 & 97.0 & 97.2 \\
    \bottomrule[0.6pt]
  \end{tabular}
  \caption{Task quality. \name{} maintains quality comparable to BF16 at a $31.3$--$55.7\times$ smaller checkpoint size.}
  \label{tab:task-quality}
  \vspace{-10pt}
\end{table*}

\noindent\textbf{Workloads and Metrics.}
We evaluate task quality on LongBench~\cite{bai2024longbench} and RULER~\cite{hsieh2024ruler}, and Perplexity (PPL) on WikiText-2~\cite{merity2017pointer}. All quality benchmarks execute real prefix-hit recovery in the serving engine, reconstructing compressed checkpoints at the prefix boundary before continuing generation.
We evaluate end-to-end serving on four multi-turn and agentic workloads (Table~\ref{tab:serving-workloads}). We report TTFT, effective cache capacity, and token reuse ratio for prefix reuse (\S\ref{sec:eval-end-to-end}), and TTFT, end-to-end latency, and TPOT for encoding overhead (\S\ref{sec:eval-overhead}).

\vspace{-10pt}
\subsection{Compression Effectiveness}
\label{sec:eval-compression}

\noindent\textbf{Task Quality.}
Table~\ref{tab:task-quality} compares \name{} against BF16 and INT8 across downstream tasks.
\name{} generalizes across model families and scales, covering scalar-decay GDN (Qwen3.5-9B, Qwen3.8-27B) and vector-decay KDA MoE (Kimi-Linear).
Qwen3.8-27B achieves perfect scores on every RULER setting, and the smaller 9B model closely matches BF16, showing accuracy does not rely on over-parameterization.

Across task types, \name{} preserves fidelity on both synthetic recall and real-world document benchmarks.
On RULER, accuracy stays stable from 8K to 32K tokens with only occasional 1 to 2 point dips, confirming that closed-loop coding prevents error accumulation along long chains.
On LongBench, 13 of 15 scores stay within one point of BF16, matching or beating INT8 at over an order of magnitude less memory.
The two remaining drops are 1.6 points on Kimi-Linear NarrativeQA (still ahead of INT8) and 1.2 points on Qwen3.8-27B RepoBench-P, offset by comparable gains elsewhere.

\begin{figure}[t]
  \centering
  \includegraphics[width=0.95\columnwidth]{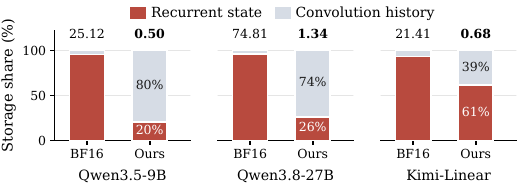}
  \vspace{-10pt}
  \caption{Checkpoint storage breakdown.
  Numbers above bars give average total storage per checkpoint in MiB.}
  \label{fig:checkpoint-storage}
  \vspace{-10pt}
\end{figure}

\noindent\textbf{Checkpoint Storage.}
Figure~\ref{fig:checkpoint-storage} compares the average per-checkpoint storage of BF16 and \name{}.
On recurrent states alone, \name{} shrinks storage by $237.5\times$ on Qwen3.5-9B, $204.6\times$ on Qwen3.8-27B, and $47.7\times$ on Kimi-Linear.
Because recurrent states become so small, short-convolution history replaces them as the dominant component on Qwen, accounting for 74\% to 80\% of compressed bytes compared with under 7\% in BF16.
By storing compact pre-projection inputs and re-projecting them upon reconstruction (\S\ref{sec:prefix-hit-path}), \name{} cuts convolution history by $2.8\times$ on both Qwen models and $5.3\times$ on Kimi-Linear with negligible recomputation overhead, bringing overall per-checkpoint compression to $50.5\times$ on 9B, $55.7\times$ on 27B, and $31.3\times$ on Kimi-Linear.
In contrast, Kimi-Linear remains state-dominated (61\%) because its KDA layers carry a larger recurrent-state share.

\begin{figure}[t]
  \centering
  \includegraphics[width=0.95\columnwidth]{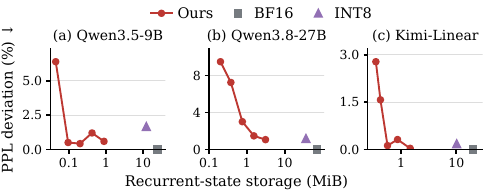}
  \vspace{-10pt}
  \caption{PPL deviation versus per-checkpoint state storage.}
  \label{fig:quality-storage}
  \vspace{-10pt}
\end{figure}

\noindent\textbf{Quality and Space Trade-off.}
Figure~\ref{fig:quality-storage} shows perplexity deviation from BF16 across per-checkpoint storage budgets, with five \name{} points doubling per-head rank budgets from $0.5\times$ to $8\times$.
Because uniform INT8 retains full-rank matrices with uniform precision, it cuts state storage by only $2\times$.
In contrast, \name{} exploits inter-state redundancy through chained low-rank residuals, shifting the trade-off curve left by over an order of magnitude across all three models.
Beyond the tightest budget, perplexity deviation drops sharply and converges near BF16, achieving lower deviation than INT8 on every model at a fraction of its storage.

\begin{figure*}[!t]
  \centering
  \includegraphics[width=0.85\textwidth]{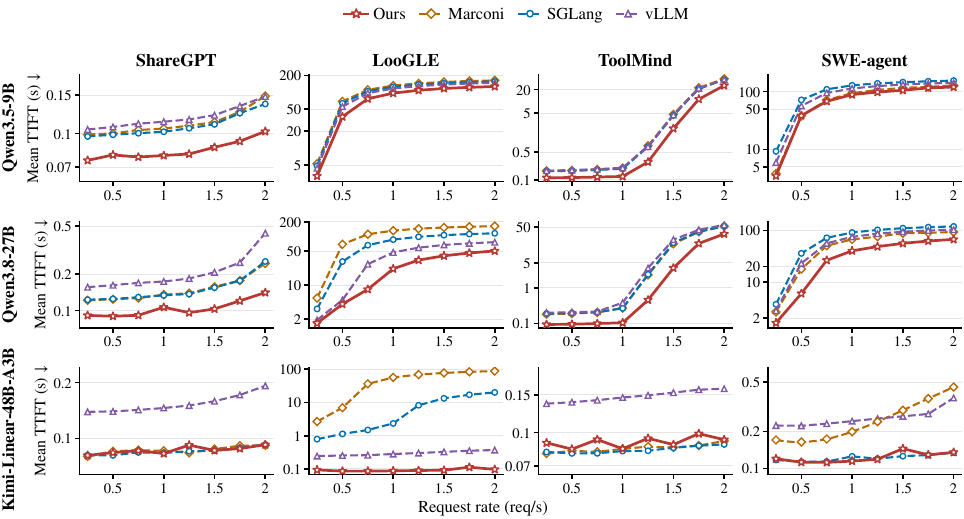}
  \vspace{-10pt}
  \caption{Mean TTFT versus request rate across three models and four workloads (log-scale vertical axes).}
  \label{fig:serving-ttft}
  \vspace{-15pt}
\end{figure*}

\begin{figure}[!t]
  \centering
  \includegraphics[width=\columnwidth]{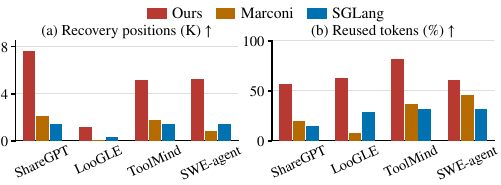}
  \vspace{-15pt}

  \caption{Recovery positions and token reuse across four workloads on Qwen3.8-27B.}
  \label{fig:effective-capacity}
  \vspace{-10pt}

\end{figure}

\vspace{-10pt}
\subsection{Serving Performance}
\label{sec:eval-end-to-end}

\noindent\textbf{Prefix-Reuse Latency.}
Figure~\ref{fig:serving-ttft} compares mean TTFT across all three models and four workloads, with all systems serving from the same populated cache snapshot while \name{} runs full state reconstruction on the critical path.
\name{} achieves the lowest TTFT whenever state capacity or checkpoint coverage limits reuse.
On Qwen3.8-27B, \name{} outperforms the fastest baseline by up to $1.75\times$ on ShareGPT, $3.30\times$ on LooGLE, $5.03\times$ on ToolMind, and $2.97\times$ on SWE-agent.
On Qwen3.5-9B, it similarly leads across all four workloads by up to $2.44\times$ over the fastest baseline.
On Kimi-Linear, \name{} cuts LooGLE TTFT by up to $3.81\times$ over vLLM while avoiding the queueing overload seen in SGLang and Marconi.
On the other three Kimi-Linear workloads, four GPUs already hold the active working set, so \name{} matches SGLang's sub-second latency floor.

Two structural limitations explain why Marconi underperforms across most workloads despite its unified memory pool and FLOP-aware eviction.
First, storing uncompressed states caps physical slot count once concurrent working sets exceed GPU memory on ShareGPT, ToolMind, and SWE-agent.
Second, to contain memory pressure, Marconi restricts checkpoint admission to request endpoints and already-observed branch points.
On two-visit document QA (LooGLE), the first query exposes no branch yet, so Marconi caches only the tail, misses shared document prefixes, and triggers queueing overload.
By comparison, vLLM preserves block-aligned checkpoints that recover most shared document prefixes, whereas SGLang's sparser coverage and per-path state cap miss many documents and still trigger queueing at higher rates.
By compressing states, \name{} breaks both limits and retains dense checkpoints for near-complete prefix recovery on LooGLE.

\noindent\textbf{Effective Cache Capacity.}
Figure~\ref{fig:effective-capacity} shows that these TTFT gains on Qwen3.8-27B stem from higher effective cache capacity under the same memory limit.
Across all four workloads, \name{} retains $3.6\times$ to $5.4\times$ more recovery positions than SGLang and $3.0\times$ to $12.1\times$ more than Marconi.
Whereas uncompressed systems must evict bulky states early or restrict admission, leaving resident KV prefixes without usable states, compact checkpoints let \name{} both capture intermediate reuse points beyond request endpoints and fit far more positions in GPU memory.
This raises token reuse across every workload, reaching up to 81.8\% on ToolMind, and near the queueing threshold cuts relative TTFT even further than token savings alone.

\begin{figure}[!t]
  \centering
  \includegraphics[width=\columnwidth]{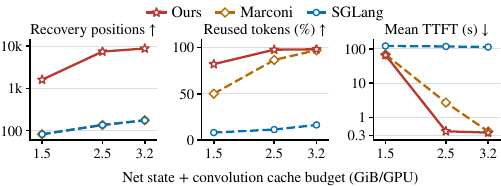}
  \vspace{-15pt}
  \caption{Recurrent state and convolution history budget sweep on Qwen3.8-27B SWE-agent (TP4).}
  \label{fig:state-budget}
  \vspace{-10pt}
\end{figure}

\noindent\textbf{State Budget Sensitivity.}
Because long-context KV caches leave only a few GiB for checkpoints, Figure~\ref{fig:state-budget} fixes the KV cache at 45\,GiB per GPU and sweeps the state and convolution budget from 1.5 to 3.2\,GiB on Qwen3.8-27B SWE-agent.
At 1.5\,GiB, \name{} already achieves much higher token reuse than Marconi and SGLang, although remaining misses still cause queueing.
At 2.5\,GiB, \name{} holds $54.8\times$ more recovery positions than both baselines, pushes token reuse to 97.5\%, and cuts mean TTFT by $6.9\times$ over Marconi while SGLang remains in queueing overload.
Marconi still outperforms SGLang at this budget, confirming that its limit comes from uncompressed state size rather than its cache policy.
\name{} approaches full trace coverage near 2.9\,GiB per GPU (mostly short-convolution history), while uncompressed baselines only match this reuse and latency at 4.5 to 6.5\,GiB.

\vspace{-10pt}
\subsection{Algorithm Ablation}
\label{sec:eval-ablation}

Figure~\ref{fig:algorithm-ablation} evaluates the impact of each design choice on WikiText-2 across all three models under the same storage budget. We compare full \name{} against three variants. Independent encoding compresses each checkpoint separately, open-loop coding removes closed-loop error feedback while keeping per-head ranks, and uniform ranks give the same rank budget to every head. Every component clearly improves both KL divergence and PPL accuracy across all models. Specifically, full \name{} reduces KL divergence by up to 64.0\% over independent encoding, 53.9\% over open-loop coding, and 38.6\% over uniform ranks.

\vspace{-10pt}
\subsection{Encoding and Decoding Efficiency}
\label{sec:eval-overhead}

\begin{figure}[!t]
  \centering
  \includegraphics[width=\columnwidth]{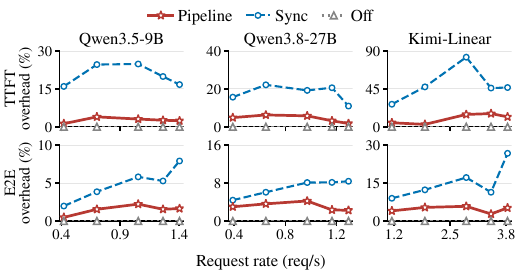}
  \vspace{-15pt}
  \caption{Mean TTFT and end-to-end latency overhead of synchronous and pipelined encoding relative to Off.}
  \label{fig:encoding-overhead}
  \vspace{-15pt}
\end{figure}

\noindent\textbf{Encoding.}
Figure~\ref{fig:encoding-overhead} compares synchronous (Sync) and pipelined (Pipeline) encoding against disabled checkpointing (Off) under cold prefill with zero prefix reuse. Compared to Sync, Pipeline sharply lowers both TTFT and end-to-end overhead across all three models, holding mean TTFT overhead within 6.3\% on Qwen and mean end-to-end overhead within 5.8\% overall. The savings mostly appear on TTFT because prefill creates most checkpoints and Sync directly stalls it. Under continuous batching, Sync's prefill stalls can also delay running decode steps, whereas Pipeline hides encoding off the forward path and keeps mean TPOT overhead within 2.8\% on Qwen and 6.0\% overall. Kimi-Linear sees slightly higher relative TTFT because its 3B-active MoE forward pass on four GPUs is much shorter, leaving a narrower window to hide encoding. Because cold prefill encodes every prompt token, these results give an upper bound on online overhead and stay far smaller than the multi-fold TTFT gains from prefix reuse.

\begin{figure*}[!t]
  \begin{minipage}[t]{\dimexpr(\textwidth-\columnsep)/2\relax}
    \centering
    \includegraphics[width=\linewidth]{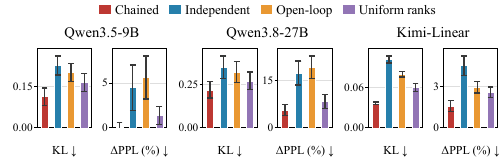}
    \vspace{-15pt}
    \caption{Algorithm ablation. $\Delta$PPL is the absolute relative PPL deviation from BF16. Error bars show one standard error.}
    \label{fig:algorithm-ablation}
  \end{minipage}\hfill
  \begin{minipage}[t]{\dimexpr(\textwidth-\columnsep)/2\relax}
    \centering
    \includegraphics[width=\linewidth]{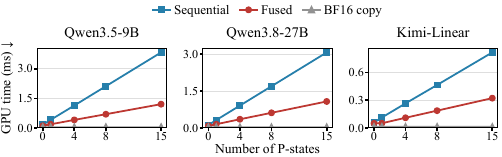}
    \vspace{-15pt}
    \caption{State reconstruction time versus chain depth.}
    \label{fig:decoding-efficiency}
  \end{minipage}
  \vspace{-15pt}
\end{figure*}

\noindent\textbf{Decoding and Reconstruction.}
Figure~\ref{fig:decoding-efficiency} compares sequential and fused reconstruction against a BF16 copy baseline for one I-state and up to 15 P-states. At 15 P-states, fusion speeds up reconstruction by $3.18\times$ on Qwen3.5-9B, $2.86\times$ on Qwen3.8-27B, and $2.54\times$ on Kimi-Linear. Full recovery, including GPU ANS decompression and short-convolution replay (median 2.38\,ms on Qwen3.8-27B TP4, including 0.28\,ms for TP all-gather), runs on the critical path of all serving experiments (\S\ref{sec:eval-end-to-end}).

\vspace{-10pt}
\section{Discussion}

\noindent\textbf{Residual Definition.}
The residual definition is not unique.
Equation~\ref{eq:gdn-state} separates scalar decay from a value-prediction correction.
Grouping decay with key-dependent erasure yields an equivalent affine state transition
\begin{equation*}\footnotesize
S_t=\alpha_t(I-\beta_t k_tk_t^\top)S_{t-1}+\beta_t k_tv_t^\top.
\end{equation*}
Across an interval, this expands to $S_t=\Phi_{s,t}S_s+\widetilde R_{s,t}$, where $\Phi_{s,t}$ is the accumulated transition and $\widetilde R_{s,t}$ accumulates writes transformed by later steps.
Because multiplying the $k_t k_t^\top$ terms produces a dense $d_k \times d_k$ matrix, storing $\Phi_{s,t}$ directly costs $d_k^2$ values per head, whereas reconstructing it step by step requires retaining intermediate keys and gates.
Instead, \name{} uses $R_{s,t}=S_t-\gamma_{s,t} S_s$, where $\gamma_{s,t}$ tracks only accumulated decay.
This choice stores only one decay scalar per head and interval for GDN, or one per key dimension for KDA, and applies uniformly across both direct key-value updates in GLA and value-prediction corrections in GDN and KDA.

\noindent\textbf{CPU Offloading.}
A natural extension is evicting checkpoints from GPU to CPU memory (\S\ref{sec:cache-mgmt}) instead of discarding them.
Because the radix tree separates metadata from record storage, cold I/P records and convolution histories can move to CPU memory while preserving chain handles, reloading only on a prefix hit.
On Kimi-Linear (TP4), transferring a 16K-token KV cache over PCIe takes 10.741\,ms, whereas moving an entire 16-checkpoint chain takes only 0.353\,ms (under 0.05\,ms per checkpoint on average).
This transfer latency is negligible, making offloading an orthogonal and promising deployment option for higher prefix hit rates.
We keep our current design GPU-resident to stay focused on compression and ensure a clean comparison under equal GPU memory budgets.

\noindent\textbf{Codec Offload.}
Although pipelined encoding keeps online latency overhead low, co-locating compression with inference still consumes GPU SM resources.
Offloading state encoding to dedicated GPUs removes contention on serving devices but adds hardware and state-transfer costs.
Alternatively, NVIDIA's NVENC/NVDEC can handle compression and decompression~\cite{xu2025llm265,mi2026kvcodec}, while SmartNICs can fetch and decompress KV caches~\cite{xiang2025shadowserve}, freeing GPU compute for inference. However, these directions remain early, whereas \name{}'s GPU kernels run portably on standard datacenter GPUs.

\vspace{-10pt}
\section{Related Work}
\noindent \textbf{Linear Attention.}
Linear sequence models replace quadratic attention with fixed-size recurrent states to bound long-context cost, spanning kernelized attention, state-space models, gated delta-rule variants, and hybrid LLMs~\cite{katharopoulos2020transformers,xiao2026ramnetlineartimesequencemodeling,gu2024mamba,dao2024mamba2,yang2024gla,yang2024deltanet,yang2025gated,qwen38_27b,kimi_linear}.
However, prefix caching must still store many bulky checkpoints across sessions and branches, which \name{} addresses through chained state compression.

\noindent \textbf{KV Cache and Recurrent State Compression.}
Prior KV compression follows three main directions, including token eviction and selective prefetching~\cite{zhang2023h2o,xiao2024streamingllm,InfiniGen}, low-bit quantization~\cite{liu2024kivi,hooper2024kvquant,kang2024gear,zhang2025diffkv,liu2025comet,zhao2024atom}, and entropy or codec-based packing for storage and transfer~\cite{liu2024cachegen,staniszewski2026kvtc,xu2025llm265,mi2026kvcodec}.
Recurrent states differ fundamentally because each checkpoint folds an entire prefix into fixed-size matrices rather than token-indexed vectors.
Recent methods quantize recurrent states during decoding~\cite{pan2026leapquant,yao2026stepquant}.
\name{} instead focuses on prefix caching, exploiting cross-checkpoint temporal redundancy through decay-based prediction and head-specific low-rank residuals.

\noindent \textbf{Prefix Caching.}
Serving engines share exact prefixes through GPU paged or radix-tree managers~\cite{kwon2023vllm,zheng2024sglang}, expand capacity across tiered host and distributed memory~\cite{gao2024cachedattention,qin2025mooncake}, and compose cached KV chunks~\cite{yao2025cacheblend,wang2026cacherepairlearningrepaircrosschunk}.
Hybrid models additionally require a matching recurrent state at every recovery position, which Marconi manages through selective admission and FLOP-aware eviction~\cite{pan2025marconi}.
Orthogonal to eviction policies~\cite{ouyang2026unicache} and tiering, \name{} compresses checkpoint states so far more recovery positions remain GPU-resident.

\vspace{-10pt}
\section{Conclusion}

We presented \name{}, a compact prefix-caching system for hybrid LLMs. \name{} organizes recurrent checkpoints into I/P-state chains with per-head rank allocation, paired with asynchronous encoding, fused reconstruction, and dependency-aware eviction. \name{} shrinks checkpoints by $31.3\times$ to $55.7\times$ and cuts mean TTFT by up to $5.03\times$ with BF16-comparable quality.

\clearpage
\nocite{ding2026tilus}
\bibliographystyle{plain}
\bibliography{references}
\end{document}